\documentclass[twocolumn,pre,showpacs]{revtex4-1}
\usepackage{graphicx}

\begin{document}
\title{Multicritical points in the three-dimensional 
  XXZ antiferromagnet with single-ion anisotropy}
\author{Walter Selke}
\affiliation{Institut f\"ur Theoretische Physik, RWTH Aachen
  University, 52056 Aachen, Germany}

\begin{abstract}
The classical Heisenberg antiferromagnet with uniaxial exchange
anisotropy, the XXZ model, and competing planar single-ion anisotropy in 
a magnetic field on a simple cubic lattice is
studied with the help of extensive
Monte Carlo simulations. The biconical (supersolid) phase, bordering
the antiferromagnetic and spin-flop phases, is found to
become thermally unstable well below the onset of the disordered, paramagnetic
phase, leading to interesting multicritical points.
\end{abstract}


\maketitle
Uniaxially anisotropic Heisenberg antiferromagnets in a
magnetic field have been
studied quite extensively in the past, both experimentally
and theoretically \cite{rev1,rev2}. Usually, they display, at low
temperatures and fields, the antiferromagnetic (AF) phase and, when increasing
the field, the spin--flop (SF) phase. A prototypical model
describing these phases as well as multicritical
points, is the Heisenberg model with a
uniaxial exchange anisotropy, the XXZ model
 \begin{equation}
  {\cal H}_{xxz} = J \sum\limits_{i,j}
  \left[ \, \Delta (S_i^x S_j^x + S_i^y S_j^y) + S_i^z S_j^z \, \right]
  \; - \; H \sum\limits_{i} S_i^z
\end{equation}
where $J > 0$ is the antiferromagnetic exchange coupling between classical
spins, $(S_{i(j)}^x,S_{i(j)}^y,S_{i(j)}^z)$, of length one
at neighboring sites, $i$ and
$j$, on a simple cubic lattice, $\Delta$ is the uniaxial exchange
anisotropy, $1 > \Delta > 0$, and $H$ is the applied
magnetic field along the easy axis, the $z$--axis.- Recently, we
presented numerical evidence for the existence of a bicritical
point in the Heisenberg universality class \cite{selke}, at which
the AF, SF and paramagnetic (P)
phases meet. Especially, there the value of the critical isotropic
Binder cumulant agrees closely with the one of the perfect
Heisenberg model.  The scenario is in accordance with early
renormalization group (RG) analyzes
in one--loop--order \cite{FN,KNF}, as well as a subsequent RG calculation
in two--loop--order
\cite{Folk} showing that such a bicritical point may exist. Note
that a conflicting characterization of the multicritical AF-SF-P point as
a 'biconical fixed point'\cite{CPV,HV} has
been suggested, mainly based on high-order RG arguments, augmented by
Monte  Carlo simulations.

In this contribution, we shall consider the XXZ model, Eq. (1), with
an additional planar single-ion anisotropy, competing with the
uniaxial exchange anisotropy. This term reads  

 \begin{equation}
  {\cal H}_{si} = D \sum\limits_{i} (S_i^z)^2
\end{equation}

The full model, ${\cal H}_{xxz} +{\cal H}_{si}$, has been investigated, in the
last few years, carefully, especially, for $S= 1$ quantum
antiferromagnetic spin chains, displaying a supersolid
phase \cite{BS,PCS,Faz}, being the
analog of the mixed or biconical (BC) phase \cite{MT,KNF,Nu} in standard
notation for magnets. One may also mention a study for the classical full
model on square lattices \cite{HS}. Again, the BC phase arises from
the competition between the uniaxial exchange and planar
single-ion anistropies. 

In the BC phase, see Fig.1, spins on the bipartite
sublattices form cones with different opening angles so that
one gets coexisting orders of the staggered transverse, $xy$, 
spin components, like in the SF phase, as well as longitudinal, $z$,
components, like in the AF phase. 

\begin{figure}
\resizebox{0.95\columnwidth}{!}{%
  \includegraphics{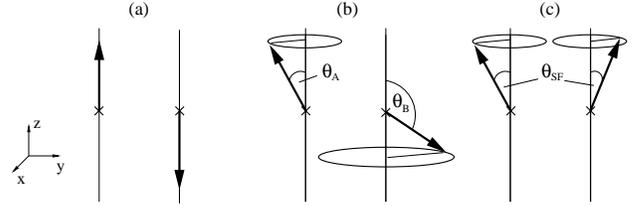}
}
\caption{Spin orientations on neighboring sites showing
     antiferromagnetic (a), biconical (b), and spin--flop (c) ground
     state configurations. }
\label{fig:1}
\end{figure}

The resulting ground state phase diagram for the classical model
on a simple cubic lattice may be determined
analytically \cite{MT,HS}. Possible ground states
are depicted in Fig. 1, showing AF, SF, and BC spin configurations.
A concrete example of the ground state phase diagram in
the $(H/J, D/J)$  plane, setting  $\Delta =0.8$, is displayed
in Fig. 2. The BC structure is seen 
to intervene between
the AF and SF structures for non-zero planar single-ion anisotropy
$D/J < 0.6$. The opening angles $\Theta_A$ and $\Theta_B$
are uniquely related, depending on the field, continuously approaching
the limiting values of the AF and SF orientations. The BC structure
reaches its maximal extent close to $D/J = 0.4$. The lower
critical field between the AF and BC ground
states, $H_{ab}$, is given by

\begin{equation}
  H_{ab} =  \sqrt{ (6J-2D)^2 - (6J\Delta)^2 }
\end{equation}

\noindent
and the upper critical field, separating the BC
and SF ground states, $H_{bs}$, is given by

\begin{equation}
  H_{bs} =  \left[ \, 36J^2 - (6J\Delta + 2D)^2 \, \right] / H_{ab}.
\end{equation}

\noindent
The transition from the SF to the ferromagnetic (F) ground state occurs
at the field $H_{sf}$ with

\begin{equation}
  H_{sf} =  6J(1+ \Delta) +2D.
\end{equation}

At non-vanishing temperatures, the BC phase may be expected
to lead to a tetracritical point \cite{Aha}, at which the
AF, SF, BC, and P phases meet. One of the aims of the 
present study has been to clarify this expectation by using
extensively Monte Carlo (MC) techniques \cite{LanBinMC}.

Following previous MC studies on the XXZ model and
variants \cite{selke,HS,LanBin,Ban}, we fixed the 
exchange anisotropy at $\Delta =0.8$. The single-ion anisotropy
has been chosen to be, mostly, $D/J =0.2$. At the end of this
paper, results for
$D/J =0.4$ will be briefly discussed as well. To take into 
account systematically finite-size effects \cite{Barber}, we
simulated cubic lattices with
$L^3$ spins, with $L$ ranging from 4 to 32, employing full periodic
boundary conditions. To obtain data of the desired accuracy, runs
with, at least, $5 \times 10^7$ Monte Carlo steps
per spin were performed, applying the standard Metropolis
MC algorithm. Error bars result from averaging
over a few of such runs. When error bars are not depicted in the figures, then
they are smaller than symbol sizes.

To determine the phase diagram, we record, especially, quantities related
to the Ising, XY, Heisenberg, and biconical order parameters. In
particular, we monitor the absolute values 
of the staggered longitudinal, $m^z_{st}$, transverse,
$m^{xy}_{st}$, and isotropic, $m^{xyz}_{st}$, magnetizations as
well as of the product of the staggered longitudinal and
transverse magnetization
components. Furthermore, related staggered susceptibilities, namely, 
$\chi ^z_{st}, \chi ^{xy}_{st}$, and $\chi^{xyz}_{st}$, are computed. In
addition, we calculate corresponding
longitudinal, transverse, and isotropic Binder cumulants
\cite{Binder}, $U^z, U^{xy}$, and $U^{xyz}$.
\begin{figure}
\resizebox{0.85\columnwidth}{!}{%
  \includegraphics{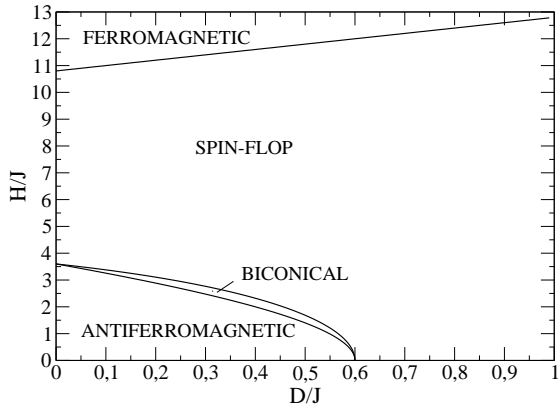}
}
  \caption{Ground states in the $(H/J,D/J)$ plane of the full
    model on a cubic lattice
    with uniaxial exchange anisotropy $\Delta= 0.8$ and planar
    single-ion anisotropy, $D/J > 0$.}
\label{fig:2}
\end{figure}

\vspace{0.5cm}

\begin{figure}
\resizebox{0.88\columnwidth}{!}{%
  \includegraphics{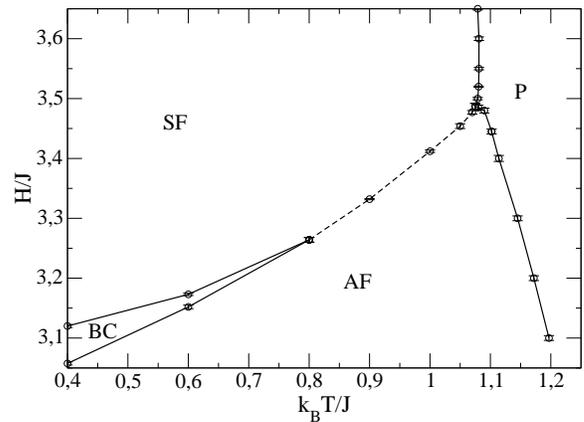}
}
\caption{Phase diagram in the temperature-field plane
of the three-dimensional full model 
with $\Delta =0.8$ and planar single-ion anisotropy $D/J =0.2$. The broken
line indicates transitions of first order.}
\label{fig:3}
\end{figure}
\vspace{1.5cm}

\vspace{-1.4cm}

The interesting part of the resulting phase diagram is depicted
in Fig. 3: Two distinct multicritical points occur, the AF-SP-P
and the AF-BC-SF points.

As before \cite{selke}, to obtain
the AF--P boundary, with the transition being in the
Ising universality class, the size dependence of the height
and position of the maximum in the longitudinal staggered
susceptibility, $\chi^z_{st}$, allows for reliable
estimates. $U^{xy}$ turns out to be very useful in estimating the SF--P
transition line. There one observes 
rather small finite--size correction terms to the critical Binder
cumulant for cubic lattices in the XY universality
class \cite{hase1}, $(U^{xy})^*$= 0.586.., with the critical
Binder cumulant being the cumulant at the transition in
the thermodynamic limit, $L \rightarrow \infty$. It
turned out to be less straightforward to identify the
transition between the AF and SF phases, with, possibly,
an intervening BC phase.

Indeed, the extent of the BC phase follows from the 
behavior of the staggered
longitudinal and tranverse susceptibilities, $\chi ^z_{st}$ and
$\chi ^{xy}_{st}$, with the AF-SF transition being signalled by the peak in
$\chi ^{xy}_{st}$, and the BC-SF transition being indicated
by the location of the maxima in the longitudinal
susceptibility. Obviously, the BC phase shrinks with increasing 
temperature, becoming, eventually, very narrow. Accordingly, further
analyzes are needed
to check the possible vanishing of the BC phase, which would then
imply the existence of an AF-BC-SF multicritical point.
 
A convenient way for identifying the AF-BC-SF point is 
demonstrated in Fig. 4. There, the size dependence of the height of the
maxima in the staggered longitudinal and transverse
susceptibilities is shown, both in the case of an 
intervening BC phase, at the lower temperature, $k_BT/J =0.6$, and
in case of a direct AF-SF transition of first order, at the higher
temperature, $k_BT/J= 0.9$.

In fact, at the higher temperature, the maxima of both
susceptibilties grow to a good approximation like $L^3$, for
$L$ exceeding about 8, see Fig. 4. This
behavior is in accordance with a transition of first
order. In marked contrast, at
the lower temperature, $k_BT/J =0.6$, both maxima occur at distinct
fields, reflecting the separate  AF-BC and BC-SF transitions. Moreover, the
heights of the peaks grow much weaker with system size: The effective
exponents, described
by the local slopes between successive points in the doubly
logarithmic plot, are clearly smaller than 3, as illustrated in
Fig. 4. Actually, for both susceptibilities
we find similar results for the effective exponents, dropping down, for
the largest sizes we studied, to values even well
below 2, to about 1.6. 

Actually, one may expect the AF-BC transition
to belong to the XY, and the BC-SF transition to the Ising
universality class \cite{KNF}. A detailed
description of these transitions is, however, beyond the scope of the
present study. We also refrained from attempting to
locate the AF-BC-SF point very accurately, and to elucidate
its (critical) properties. Its position is roughly at
$k_BT_{abs}/J= 0.85 \pm 0.05$.
\begin{figure}
\resizebox{0.85\columnwidth}{!}{%
  \includegraphics{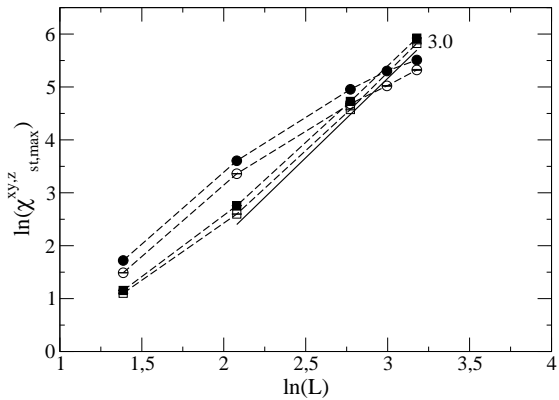}
}
  \caption{Finite size behavior of the maximal staggered
longitudinal, $z$ (open symbols),
and transverse, $xy$ (full symbols), susceptibilities for linear
dimension, $L$, ranging from 4 to 24, at $k_BT/J =0.6$ (circles) and
0.9 (squares), with $\Delta= 0.8$ and $D/J= 0.2$. For
comparison, the dependence $\propto L^3$, characteristic for a transition
of first order, is shown as solid line.}
\label{fig:4}
\vspace{0.8cm}
\end{figure}
\begin{figure}
\resizebox{0.85\columnwidth}{!}{%
  \includegraphics{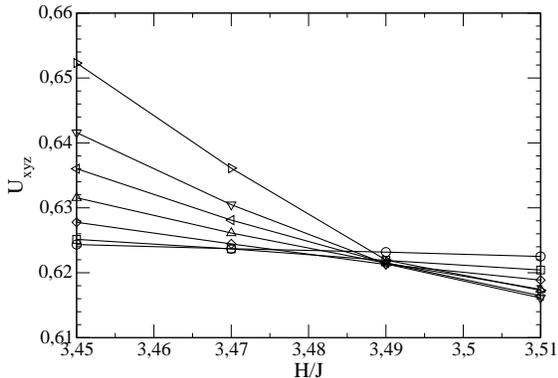}
}
\caption{Isotropic Binder cumulant $U^{xyz}$, for $\Delta =0.8$
and $D/J= 0.2$ at $k_BT/J= 1.075$, varying
the field, for lattice sizes $L= 4$ (circles), 8 (squares), 12 (diamonds),
16 (triangles up), 20 (triangles left), 24 (triangles down),  and
32 (triangles right). }
\label{fig:5}
\end{figure}

In any event, the AF-BC-SF point occurs well below the
AF-SF-P multicritical
point, $k_BT_{asp}/J= 1.075 \pm 0.01$, $H_{asp}/J= 3.490 \pm 0.002$, at 
which the continuous
AF-P and SF-P phase transition lines meet the
AF-SF transition line of first order. Obviously, the
upper multicritical point reminds of the multicritical point
in the XXZ model on the cubic lattice. Accordingly, the planar
single-ion anisotropy leads to a BC phase, but it does not lead to a
tetracritical AF-BC-SF-P point, at least, in the case we
considered so far, $D/J =0.2$.

To elucidate the nature of the AF-SF-P multicritical point, we
studied in its vicinity, especially, the isotropic
Binder cumulant $U^{xyz}(T,H,L)$. In case of a bicritical
point with Heisenberg symmetry,
the critical Binder cumulant $(U^{xyz})^*$ may be expected
\cite{hase2,pec} to acquire the characteristic 
value 0.620 ($\pm 0.003$), using
periodic boundary conditions for lattices of cubic shape.

Similarly to the findings for
the perfect XXZ model, $D=0$, we may distinguish three different
scenarios, when varying the temperature and monitoring the
size dependence of $U^{xyz}$ as a function of the field $H$ \cite{selke}.

Below the AF-SF-P point, e. g. , at $k_BT/J= 1.05$, $U^{xyz}$ increases
with larger lattice size $L$, so that there are no intersection points
of the isotropic cumulants near the direct AF-SF transition.

At $k_BT/J= 1.075$, see Fig. 5, all
intersection points of the isotropic cumulants, for lattices of
sizes $L$ ranging from 8 to 32, occur closely to the
characteristic Heisenberg value,  $(U^{xyz})^* \approx 0.620$. This
fact may be interpreted as evidence for being in the immediate 
vicinity of a bicritical point of Heisenberg symmetry. Actually,
moving to higher temperatures, another scenario shows up. For
example, at $k_BT/J= 1.1$, there
are still intersection points of the cumulants for different
lattice sizes, but they shift systematically for larger lattices
to lower and lower values below that of the critical cumulant in
the Heisenberg case. For instance, the intersection 
point of the cumulants for the two largest sizes, $L= 24$ and 32, is
even below 0.58.

As for the perfect XXZ model, these
observations on the isotropic Binder cumulant are
consistent with the AF-SF-P point, at $D/J= 0.2$, being a
bicritical point in the Heisenberg universality class.

Note that the staggered isotropic susceptibility seems to be 
the less sensitive quantity to identify the nature of the 
AF-SF-P point, with the values of the critical exponent for
bicritical and biconical multicritical points being very
close to each other \cite{HV}.

Finally, we mention briefly results of our (preliminary) simulations
for an enhanced planar single-ion anisotropy, $D/J= 0.4$. Again, there
is no evidence for a tetracritical AF-BC-SF-P point. Instead, we find an
AF-BC-SF point. It is followed, at higher temperatures, by
a direct transiton of first order between the AF and SF
phases, with the BC phase being squeezed out. Upon further 
increase of the temperature, an AF-SF-P multicritical point occurs.

I should like to thank R. Folk and M. Hasenbusch for useful discussions.


\begin{thebibliography}{}

\bibitem{rev1}  Y.\ Shapira, in {\it Multicritical Phenomena} , ed. by
  R.\ Pynn and A. Skjeltorp (Plenum Press, New York and London, 1984),
  p.35; and references therein.
\bibitem{rev2} W.\ Selke, M.\ Holtschneider, R.\ Leidl, S.\ Wessel,
  G.\ Bannasch, and D.\ Peters, Physics Procedia \textbf{6}, 84
  (2010); and references therein.
\bibitem{selke} W.\ Selke, Phys.\ Rev.\ E \textbf{83}, 042102 (2011).
\bibitem{FN} M.\ E.\ Fisher and D.\ R.\ Nelson, Phys. Rev. Lett. {\textbf{32}},
    1350 (1974).
\bibitem{KNF} J.\ M.\ Kosterlitz, D.\ R.\ Nelson, and  M.\ E.\ Fisher, Phys. Rev. B \textbf{13}, 412 (1976).
\bibitem{Folk} R.\ Folk, Yu.\ Holovatch, and G.\ Moser,
  Phys.\ Rev.\ E \textbf{78}, 041124 (2008).
\bibitem{CPV} P.\ Calabrese, A.\ Pelissetto, and E.\ Vicari,
  Phys.\ Rev.\ B\ \textbf{67}, 054505 (2003).
\bibitem{HV} M.\ Hasenbusch and E.\ Vicari, Phys.\ Rev.\ B\ \textbf{84}, 125136 (2011).
\bibitem{BS} P.\ Sengupta and C.\ D.\ Batista, Phys.\ Rev.\ Lett\ \textbf{99}, 217205 (2007).
\bibitem{PCS} D.\ Peters, I.\ P.\ McCulloch, and W.\ Selke,
  Phys.\ Rev.\ B\ \textbf{79}, 132406 (2009); Phys.\ Rev.\ B\ \textbf{85}, 054423 (2012).
\bibitem{Faz} D.\ Rossini, V.\ Giovannetti, and R.\ Fazio,
  Phys.\ Rev.\ B\ \textbf{83}, 140411(R) (2011).
\bibitem{MT} H.\ Matsuda and T.\ Tsuneto, Prog.\ Theoret.\ Phys.\ Suppl.\
\textbf{46}, 411 (1970).
\bibitem{Nu} Z.\ Nussinov, Physics \textbf{1}, 40 (2008).
\bibitem{HS} M.\ Holtschneider and W.\ Selke, Eur.\ Phys.\ J.\ B\ \textbf{62},
  147 (2008).
\bibitem{Aha} A.\ Aharony, J.\ Stat.\ Phys.\ \textbf{110}, 659 (2003).
\bibitem{LanBinMC} D.\ P.\ Landau and K.\ Binder, {\it A Guide to
    Monte Carlo  Simulations in Statistical Physics} (University
  Press, Cambridge, 2005).
\bibitem{LanBin} D.\ P.\ Landau and K.\ Binder, Phys.\ Rev.\ B\
  \textbf{17}, 2328 (1978).
\bibitem{Ban} G.\ Bannasch and W.\ Selke,
  Eur.\ Phys.\ J.\ B\ \textbf{69}, 439 (2009).
\bibitem{Barber} M.\ N.\ Barber, in {\it Phase Transitions and
    Critical Phenomena}, ed. by C.\ Domb and J.\ L.\ Lebowitz (Academic
  Press, New York, 1983), Vol. 8.
\bibitem{Binder} K.\ Binder, Z. Physik B- Cond. Matt. \textbf{43}, 119 (1981).
\bibitem{hase1} M.\ Hasenbusch and T.\ T\"or\"ok, J. Phys. A-
  Math. Gen. \textbf{32}, 6361 (1999).
\bibitem{hase2} M.\ Hasenbusch, J. Phys. A- Math. Gen. \textbf{34}, 8221 (2001).
\bibitem{pec} P.\ Peczak, A.\ M.\ Ferrenberg, and D.\ P.\ Landau,
  Phys.\ Rev.\ B\ \textbf{43}, 6087 (1991).
\end{thebibliography}
\end{document}